\documentclass[preprint,showpacs,showkeys]{revtex4}
\usepackage[T2A]{fontenc}
\usepackage[cp1251]{inputenc}
\usepackage[english]{babel}
\usepackage{amsmath}
\usepackage{amsfonts}
\usepackage{graphicx}
\usepackage{epsfig}
\usepackage{subfigure}
\usepackage{amssymb}
\usepackage[indent,small]{caption2}

\begin{document}

\newcommand{\hdblarrow}{H\makebox[0.9ex][l]{$\downdownarrows$}-}
\title{Electron-Hole Liquid in the Couple Quantum Wells}

\author{V.S. Babichenko$^1$, I.Ya. Polishchuk$^{1,2}$, A.I. Pavlov$^1$, A. Guseynov$^1$ and M.I. Gozman$^1$}

\affiliation{1: Kurchatov Institute, Kurchatov Sq.1, 123182, Moscow, Russia\\
Tel.:+7-499-1967584\\ Fax:+7-499-9430074\\
\email{vsbabichenko@hotmail.com}
\\2: Moscow Institute of Physics and Technology,
              Theoretical department, 141700, Moscow Region,
              Dolgoprudniy, Institutskii per. 9\\ \email{iyppolishchuk@gmail.com}}


\begin{abstract}

It is shown that the homogeneous state of the spatially separated
electrons and holes in the coupled quantum wells (CQW) is instable
if the layer charge density is smaller than the critical value
$n_{c}$ specified by the parameters of the CQW. The effect is due
to the many-body Coulomb correlations which provide the
\textit{positive} compressibility. The instability results in the
formation of the inhomogeneous system which comprises the liquid
electron-hole drops.

\end{abstract}



\maketitle

\section{Introduction}

The investigation of the spatially separated electrons and holes in the
coupled quantum wells (CQW) is initially motivated by expecting that the
electron-hole pairs forming a long-living bound state, namely exciton, can
experience the Bose-Einstein condensation \cite{Lozovik}. In such systems
the electrons are located within one layer of the CQW while the holes, which
are spatially separated from the electrons, are located in the other layer.
The interest in the CQW has greatly grown in the recent years due to the
increasing ability to manufacture the high quality quantum well structures
in which electrons and holes are confined in the different spatial regions
between which the tunneling can be made negligible \cite{Review2011}. As
early as decade and a half the existence of an electron-hole condensate
phase in the CQW was predicted using the variational approach \cite%
{Lozovik-Berman1996}.

Recently, the effect of the many-body Coulomb correlations in the CQW on the
ground state of the electron-hole system is investigated in Ref. \cite%
{JETLET}. The layers of the CQW are supposed so thin that the in-layer
charge motion is a 2D one. In paper\cite{JETLET}, the compressibility is
found to be \textit{positive} if the initial layer charge density $n<n_{c},$
the critical value $n_{c}$ being specified by the physical parameters of the
CQW. The positive compressibility means the instability resulting in the
formation of the inhomogeneous system which comprises the liquid
electron-hole drops. Also, it was found in Ref.\cite{JETLET} that the
homogeneous exciton gas phase possesses the higher energy as compared to the
inhomogeneous state (of the same average density) which involves the
electron-hole liquid drops. Alternative scenarios for the formation of the
condensed phase in the electron-hole system in the CQW are also considered
in various papers, in particular in Refs. \cite{sugakov, Wilkes, Kuznetzova}.
Note that the electron-hole liquid of the same origin as in Ref.\cite%
{JETLET} was predicted for the conventional 3D-semiconductors in \cite%
{keldysh1, rice}.

The results obtained in Ref.\cite{JETLET} are based on the
assumption that each of the layers of the CQW is a many-valley
semiconductor. Thus, the every kind of the electrons or the holes
is specified by the number of the valley and the spin projection.
Let the number of different kind of the electrons as well as the
number of different kind of the holes $\nu \gg 1$. Two limiting
situations are considered in this paper: the inter-layer
separation $l\ll a_{B}/\nu $ and $l\gg a_{B}/\nu$, $a_{B}$ being
the effective Bohr radius. In the first case, the electron-hole
liquid drops are formed which possess the in-layer density
$n_{c}\sim \nu ^{3/2}$ . In the second case, $n_{c}\sim
1/l^{3/2}.$ These results are obtained using the diagrammatic
approach and the following approximation is used. For any order in
the Coulomb interaction, only the diagrams are held which are of
the minimal order in the small parameter $1/\nu .$ In this case,
the set of the
diagram is that of the RPA kind. A similar approach was developed in Ref.%
\cite{babich} to justify the formation of the electron-hole liquid drops in
conventional 3D-semiconductors. Note that the standard RPA is justified for
the high-density electron-hole plasma. In Refs \cite{babich,JETLET} the RPA\
diagram approach is based only on the assumption that the electron-hole
plasma is many-component.

In the current paper we predict the instability of the uniform electron-hole
system in the CQW for the case $l\ll a_{B}/\nu $ in the approximation next
beyond to the RPA. For this purpose, we investigate the pole of the
4-fermion vertex function in the approximation which along with the RPA-kind
diagrams takes into account the diagrams of the next order in the parameter $%
1/\nu .$ It is shown that the instability, connected with the pole of the
vertex function, corresponds to the thermodynamical instability induced by
the positive compressibility revealed in Ref. \cite{JETLET}.

Strictly speaking, the result obtained is justified for the
many-component system, $\nu \gg 1.$ However, comparing the result found in
Ref.\cite{JETLET} (obtained in the main $1/\nu $ approximation) with that
found in the current paper (obtained in the next approximation in the
parameter $1/\nu $) manifests that the critical concentration is insensitive
to the approximation. For this reason, one can expect that the instability
revealed takes place if the parameter $\nu $ is not very large.

For the sake of simplicity, the system of units is used with the effective
electron (hole) charge $e=1,$ the effective electron (hole) masses $%
m_{e}=m_{h}=1,$ and the Planck constant $\hbar =1$. Then, the effective Bohr
radius is $a_{B}=1$ and the energy is measured in the Hartree units $%
m_{e}e^{4}/\hbar ^{2}=1$.

\section{The Hamiltonian and the Vertex Function}

The Hamiltonian of the system is $\widehat{H}=\widehat{H}_{0}+\widehat{V},~%
\widehat{H}_{0}$ being the kinetic energy and $\widehat{V}$ being the
Coulomb interaction. In the second quantization one has 
\begin{equation}
\begin{array}{l}
\widehat{H}_{0}=\sum\limits_{\alpha \sigma \mathbf{k}}\frac{k^{2}}{2}%
a_{\alpha \sigma }^{+}\left( \mathbf{k}\right) a_{\alpha \sigma }\left(
\mathbf{k}\right) ; \\
\widehat{V}=\frac{1}{2S}\sum\limits_{\alpha \alpha ^{\prime }\sigma \sigma
^{\prime }\mathbf{kk}^{\prime }\mathbf{q}}V_{\alpha \alpha ^{\prime }}\left(
\left\vert \mathbf{q}\right\vert \right) a_{\alpha \sigma }^{+}\left(
\mathbf{k}\right) a_{\alpha ^{\prime }\sigma ^{\prime }}^{+}\left( \mathbf{k}%
^{\prime }\right) a_{\alpha ^{\prime }\sigma ^{\prime }}\left( \mathbf{k}%
^{\prime }-\mathbf{q}\right) a_{\alpha \sigma }\left( \mathbf{k}+\mathbf{q}%
\right) .%
\end{array}
\label{eq1}
\end{equation}%
Here $\alpha =e$ stands for the electrons, while $\alpha =h$ stands for the
holes, $\sigma =1,...,\nu $ labels the kind of the electron or the hole; $%
a_{\alpha \sigma }^{+}\left( \mathbf{k}\right) $ and $a_{\alpha \sigma
}\left( \mathbf{k}\right) $ are the creation and annihilation operators, $%
\mathbf{k}$ is the in-layer $2D$ momentum, and $S$ is the area of the
layers. The Coulomb interaction in the momentum representation, which is
assumed to be independent on the valley, reads
\begin{equation}
V_{\alpha \alpha ^{\prime }}\left( \left\vert \mathbf{k}\right\vert \right)
=\left\{
\begin{array}{c}
V=\frac{2\pi }{\left\vert \mathbf{k}\right\vert },\alpha =\alpha ^{\prime };
\\
V^{\prime }=-\frac{2\pi }{\left\vert \mathbf{k}\right\vert }e^{-\left\vert
\mathbf{k}\right\vert l},\alpha \neq \alpha ^{\prime }.%
\end{array}%
\right.   \label{eq2}
\end{equation}%
The Fermi momentum and the Fermi energy for the electrons and the holes are
the same and are equal to $p_{F}=2\pi ^{1/2}\left( n/\nu \right) ^{1/2}$ and
$\varepsilon _{F}=2\pi n/\nu $, respectively, $n$ being the layer
concentration of the electrons or the holes. It is assumed that the
temperature $T\ll \varepsilon _{F}$ and the concentration $n\ll \nu ^{3}.$

Let us consider the four-fermion vertex function $\Gamma _{\alpha \sigma
,\alpha ^{\prime }\sigma ^{\prime }}=\Gamma _{\alpha \sigma ,\alpha ^{\prime
}\sigma ^{\prime }}\left( p_{1};p_{2};k\right) $ in the Matsubara
representation. The subscripts $\alpha \sigma $ and $\alpha ^{\prime }\sigma
^{\prime }$ label the two input ends. The two output ends are the same. The
notations $p_{1}=\left( \mathbf{p}_{1},i\omega _{1}\right) $ and $%
p_{2}=\left( \mathbf{p}_{2},i\omega _{2}\right) $ are used which determine
the input momentums $\mathbf{p}_{1}$ and $\mathbf{p}_{2}$ and the input
Matsubara frequencies $\omega _{1}$ and $\omega _{2}$, while the argument $%
k=\left( \mathbf{k},i\varepsilon \right) $ specifies the transfer momentum $%
\mathbf{k}_{1}$ and the transfer frequency $\varepsilon $.

\begin{figure}[h]
\centering\includegraphics[width=0.7\textwidth]{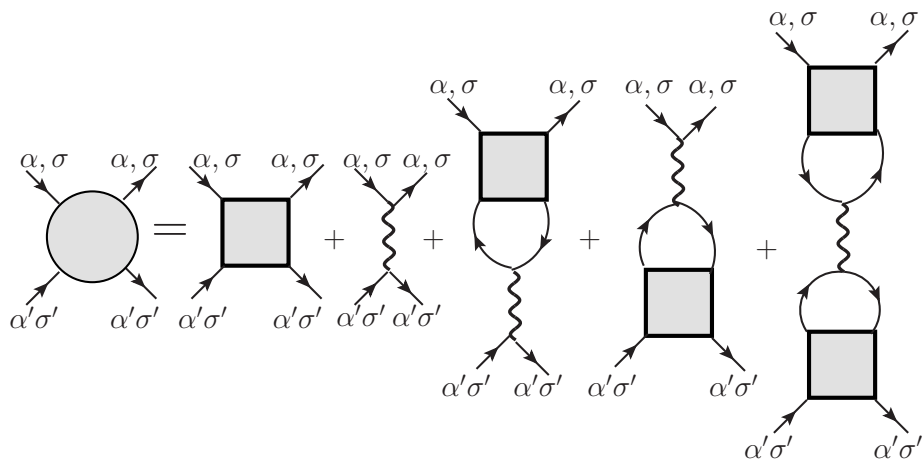}
\caption{The exact diagrammatic expansion for the vertex function $\Gamma _{%
\protect\alpha \protect\sigma ,\protect\alpha ^{\prime }\protect\sigma %
^{\prime }}$ which is represented by the gray circle. The gray squares
designate the irreducible vertex $\Gamma _{\protect\alpha \protect\sigma ,%
\protect\alpha ^{\prime }\protect\sigma ^{\prime }}^{\left( 0\right) }~$%
which is a sum of all irreducible diagrams. The wavy lines designate the
total screened Coulomb interaction $U_{\protect\alpha a^{\prime }}\left(
k\right) $ (see. Fig.\protect\ref{Figure2} ) The (internal) solid lines
designate the total Green functions $G\left( k\right) $. }
\label{Figure1}
\end{figure}

Let us introduce the concept of an irreducible diagram. A diagram
is called an irreducible one if it cannot be separated into two
parts linked only by \textit{one} interaction line $V_{\alpha
\alpha ^{\prime }}\left( \left\vert \mathbf{k}\right\vert \right)
$. Let $\Gamma _{\alpha \sigma ,\alpha ^{\prime }\sigma ^{\prime
}}^{\left( 0\right) }$ be a set of all the irreducible diagrams
which contribute into $\Gamma _{\alpha \sigma ,\alpha ^{\prime
}\sigma ^{\prime }}$. The functions $\Gamma _{\alpha \sigma
,\alpha ^{\prime }\sigma ^{\prime }}$ and $\Gamma _{\alpha \sigma
,\alpha ^{\prime }\sigma ^{\prime }}^{\left( 0\right) }$ are
connected as is shown in Fig. \ref{Figure1}

\begin{figure}[h]
\centering\includegraphics[width=100mm]{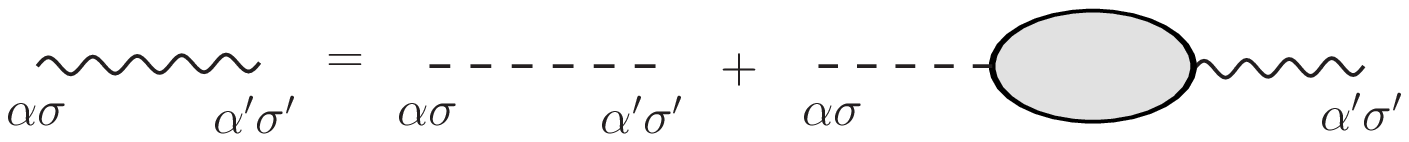}
\hfill \caption{The exact self-consistent equation for the total
screened Coulomb interaction $U_{\protect\alpha a^{\prime }}\left(
k\right) $ which is designated by the wavy line. The dashed line
represents the bare Coulomb interaction $V_{\protect\alpha
a^{\prime }}\left( \mathbf{k}\right)$. The
gray oval represents the polarization operator $\Pi _{\protect\alpha \protect%
\sigma a^{\prime }\protect\sigma ^{\prime }}\left( k\right) .$}
\label{Figure2}
\end{figure}

The wavy lines in the diagrams in Fig. \ref{Figure1} correspond to
the total screened Coulomb interaction $U_{\alpha \alpha ^{\prime
}}=U_{\alpha \alpha ^{\prime }}\left( k\right) $. This interaction
is connected with the bare interaction (\ref{eq2}) by means of the
polarization operator $\Pi _{\alpha \sigma ,\alpha ^{\prime
}\sigma }=\Pi _{\alpha \sigma ,\alpha ^{\prime }\sigma }(k)$
according to Fig. \ref{Figure2}.

\begin{figure}[h]
\includegraphics[width=0.40\textwidth]{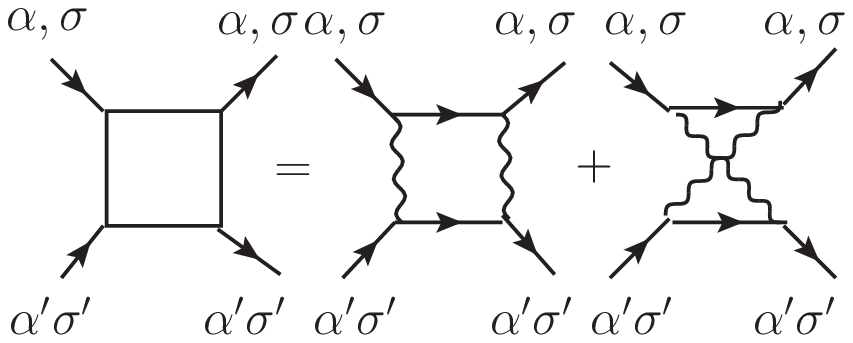}
\hfill
\includegraphics[width=0.40\textwidth]{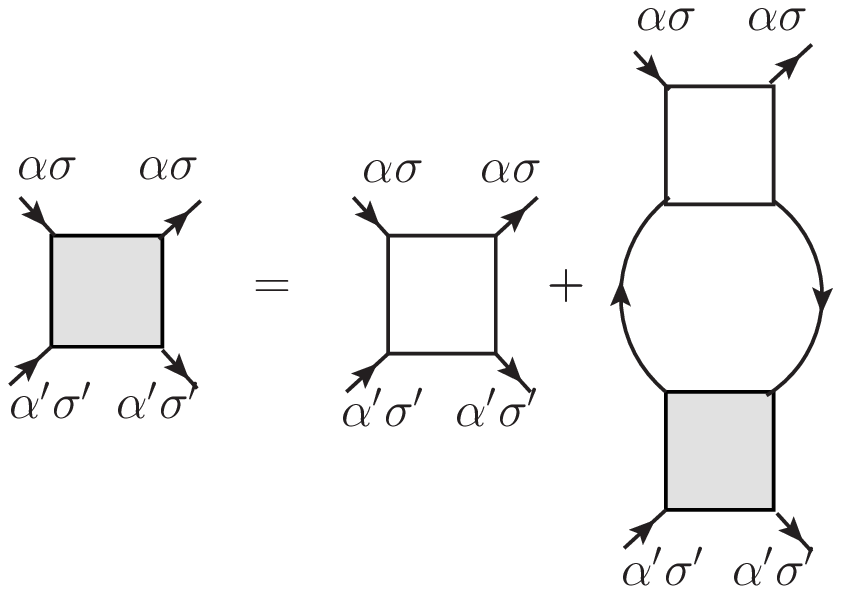}
\\
\parbox[t]{0.40\textwidth}{\caption{The white square designates the irreducible vertex
function calculated in the second order in the screened interaction $U_{\protect%
\alpha \protect\alpha ^{\prime }}$.}\label{Figure3}} %
\hfill
\parbox[t]{0.40\textwidth}{\caption{The self-consistent diagram equation
for the irreducible vertex function
$\Gamma _{\protect\alpha \protect\sigma ,\protect\alpha ^{\prime }\protect%
\sigma ^{\prime }}^{\left( 0\right) }$ main in the parameter
$\protect\nu$.}\label{Figure4.0}} %
\end{figure}

The system of the diagrammatic equations in Figs. \ref{Figure1}, \ref%
{Figure2} is not complete. To proceed, let us first turn to the irreducible
vertex $\Gamma _{\alpha \sigma ,\alpha ^{\prime }\sigma ^{\prime }}^{\left(
0\right) }.$ The minimal order in the interaction $U_{\alpha \alpha ^{\prime
}}$ for the diagrams which contribute into $\Gamma _{\alpha \sigma ,\alpha
^{\prime }\sigma ^{\prime }}^{\left( 0\right) }$ is the second order. In
this case, the vertex $\Gamma _{\alpha \sigma ,\alpha ^{\prime }\sigma
^{\prime }}^{\left( 0\right) }$ designated as $\gamma _{\alpha \sigma
,\alpha ^{\prime }\sigma ^{\prime }}$ is given by the sum of two diagrams as
provided by Fig. \ref{Figure3}

Let us consider the vertex function $\gamma _{\alpha \sigma ,\alpha ^{\prime
}\sigma ^{\prime }}$ for the momenta and frequencies which obey the
limitation
\begin{equation}
\mid \mathbf{p}_{1}\mid ,\mid \mathbf{p}_{2}\mid ,\mid \mathbf{k}\mid \leq
p_{F};\mid \omega _{1}\mid ,\mid \omega _{2}\mid ,\mid \varepsilon \mid \leq
p_{F}^{2}/2.  \label{eq10}
\end{equation}%
As is shown in the Appendix, in this case the vertex function depends
neither on the $\alpha ~$nor on the $\sigma ,$ and $\gamma _{\alpha \sigma
,\alpha ^{\prime }\sigma ^{\prime }}=-C/n^{2/3},$ $C$ being the factor of
the order of unity, $n$ being the in-layer charge concentration. Let us then
remind that the number of the valleys $\nu $ is a large parameter. For this
reason, among all the irreducible diagrams of a given order in the
interaction $U_{\alpha \alpha ^{\prime }},$ let us hold only those which are
of the minimal order in the parameter $1/\nu .$ One can convince oneself
that such sequence of the main diagrams obey the diagrammatic relation in
Fig. \ref{Figure4.0}

\begin{figure}[t]
\includegraphics[width=0.40\textwidth]{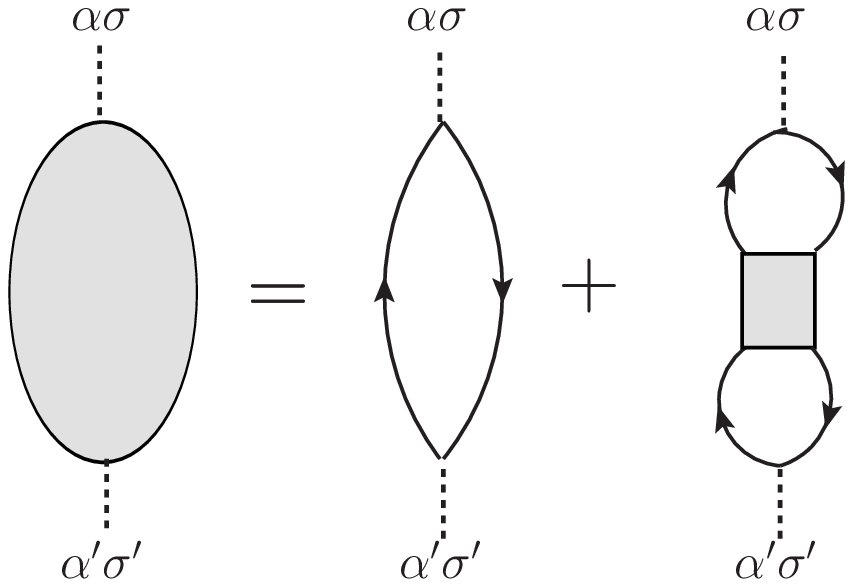}
\hfill
\includegraphics[width=0.40\textwidth]{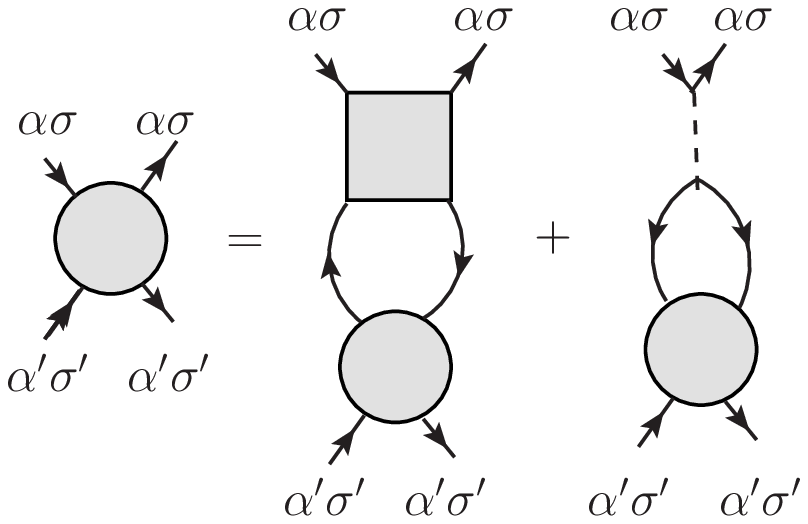}
\\
\parbox{0.40\textwidth}{ %
\caption{The polarization operator $\Pi _{\protect\alpha \protect\alpha %
^{\prime }}$ calculated in the main in the parameter $\protect\nu $
approximation. The gray square corresponds to the diagram for the $\Gamma _{%
\protect\alpha \protect\alpha ^{\prime }}^{\left( 0\right) }$
calculated according Fig. \protect\ref{Figure4.0}}
\label{Figure5.0}}
\hfill
\parbox{0.40\textwidth}{ %
\caption{The self-consistent diagrammatic relation for the vertex function $%
\Gamma _{\protect\alpha \protect\sigma ,\protect\alpha ^{\prime }\protect%
\sigma ^{\prime }}.$} %
\label{Figure6.0}}
\end{figure}

The polarization operator $\Pi _{\alpha \sigma ,\alpha ^{\prime }\sigma
^{\prime }}$, calculated in the same approximation as the vertex function $%
\Gamma _{\alpha \sigma, \alpha ^{\prime }\sigma ^{\prime}}^{\left(
0\right) }$ in Fig. \ref{Figure4.0}, is represented by the
diagrams shown in Fig. \ref{Figure5.0}.

Now, the system of equations represented by the diagrams in Figs. \ref%
{Figure1}-\ref{Figure5.0} becomes complete what allows to obtain
the self-consistent diagrammatic equation for the vertex function
$\Gamma$ shown in Fig. \ref{Figure6.0}.

Since the function $\gamma _{\alpha \sigma ,\alpha ^{\prime }\sigma ^{\prime
}}$ is a subscript independent, one can write $\Gamma =\Gamma _{e\sigma
,e\sigma ^{\prime }}=\Gamma _{h\sigma ,h\sigma ^{\prime }}$ and $\Gamma
^{\prime }=\Gamma _{e\sigma ,h\sigma ^{\prime }}=\gamma _{\alpha \sigma
,\alpha ^{\prime }\sigma ^{\prime }}.$ Then, the diagrammatic equations in
Fig. \ref{Figure6.0} reads
\begin{eqnarray}
\Gamma  &=&\left( V+\gamma \right) +\left( V+\gamma \right) \Pi _{0}\Gamma
+\left( V^{\prime }+\gamma \right) \Pi _{0}\Gamma ^{\prime },  \label{eq2+1}
\\
\Gamma ^{\prime } &=&\left( V^{\prime }+\gamma \right) +\left( V^{\prime
}+\gamma \right) \Pi _{0}\Gamma +\left( V^{\prime }+\gamma \right) \Pi_{0}\Gamma ^{\prime }.  \notag
\end{eqnarray}%
where the polarization operator{\tiny \ }$\Pi _{0}\left( p\right) $ is given
by Eq. (\ref{eq8-2}). According to Eq.(\ref{eq2}), $V^{\prime }\left(
\left\vert \mathbf{k}\right\vert \right) \approx -V\left( \left\vert \mathbf{%
k}\right\vert \right) +2\pi l$ if $\left\vert \mathbf{k}\right\vert l\ll 1.$
Then, it follows from (\ref{eq2+1}) that%
\begin{eqnarray}
\Gamma  &\approx &-\frac{1}{2\Pi _{0}}+\frac{\gamma +\pi l}{\left( 1-2\left(
\gamma +\pi l\right) \Pi _{0}\right) },  \label{eq2-1+3} \\
\Gamma ^{\prime } &=&\frac{1}{2\Pi _{0}\left( 1-2\left( \gamma +\pi l\right)
\Pi _{0}\right) }.  \label{eq2-1+4}
\end{eqnarray}%
The vertex functions (\ref{eq2-1+3}) and  (\ref{eq2-1+4}) have a pole if
\begin{equation}
1-2\left( \gamma +\pi l\right) \Pi _{0}=0.  \label{eq2-1+5}
\end{equation}%
It follows from Eq.(\ref{eq8-1-1}) and Eq.(\ref{eq8-2}) that $\Pi _{0}\left(
\left\vert \mathbf{p}\right\vert \rightarrow 0\right) =-\nu /2\pi .$ Since $%
\gamma = - C/n^{2/3}$, one obtains that the vertex functions
$\Gamma_{ee},\Gamma _{eh}$ have the pole at the concentration%
%
\begin{equation}
n_{c}=\left( \frac{C}{2\pi \left( 1/\nu +l\right) }\right) ^{3/2}.
\label{crit}
\end{equation}%
Thus, for the concentration $n<n_{c}$ the uniform state of the system is
unstable.

\section{The connection between the vertex function $\protect\gamma \left(
k\rightarrow 0\right) $ and the compressibility}

Let us give a thermodynamic interpretation of the instability point given by
Eq.(\ref{eq2-1+5}). According to paper \cite{JETLET}, within the RPA%
\begin{equation}
\mu =\frac{2\pi n}{N}+2\pi nl+\mu _{corr},~\mu _{corr}=-\frac{T}{S}%
\sum\limits_{\mathbf{p},\omega }\left( U^{\left( 0\right) }\left( p\right)
-V\left( p\right) \right) G\left( p\right) .  \label{eq.mu1}
\end{equation}%
where the function $U^{\left( 0\right) }\left( p\right) $ and the Green
function $G\left( p\right) $ are given by Eqs. (\ref{eq8-1}) and (\ref%
{eq8-1-1}). Then, one can transform this expression as follows
\[
\mu _{corr}\approx -\frac{1}{2}\int \frac{d^{2}kd\omega }{\left( 2\pi
\right) ^{3}}\left( \frac{\left[ V\left( p\right) \Pi _{0}\left( p\right) /n%
\right] }{1-2n\left[ V\left( p\right) \Pi _{0}\left( p\right) /n\right] }%
-\left( V\left( p\right) \Pi _{0}\left( p\right) /n\right) \right) .
\]%
Taking into account that the quantity $\left[ V\left( p\right) \Pi
_{0}\left( p\right) /n\right] $ does not depend on the concentration $n,$
one has%
\begin{equation}
\frac{d\mu _{corr}}{dn}=-\int \frac{d^{2}kd\omega }{\left( 2\pi \right) ^{3}}%
\left( \frac{V\left( p\right) \Pi _{0}\left( p\right) /n}{\left( 1-2nV\left(
p\right) \Pi _{0}\left( p\right) /n\right) }\right) ^{2}.  \label{eq.mu4}
\end{equation}%
Comparing this result with Eq. \ref{eq11} gives $d\mu _{corr}/dn=2\gamma .$
Then, it follows from (\ref{eq.mu1}) that%
\begin{equation}
\frac{d\mu }{dn}=\frac{2\pi }{\nu }+2\pi l+2\gamma .  \label{eq.mu5}
\end{equation}%
Thus, the instability described by the pole of the vertex function (\ref%
{eq2-1+5}) corresponds to the zero of the compressibility, which means a
thermodynamical instability.
\section{Conclusions}
It is shown that the electron-hole system in the CQW is instable if the
layer charge concentration $n<n_{c}$ (see Eq. (\ref{crit})). This
instability is determined by the pole of the vertex function. At the same
time the physical sense of this instability means just the positivity of the
compressibility. Thus, if the initial homogenouse concentration $n<n_{c}$,
the system transforms into inhomogeneouse state which contains the liquid
electron-hole drops with the density $n_{c}.$ However, if the system with
the density $n\geq n_{c}$ is created, it exists in the honogeneouse stable
state.

\section*{Appendix. The vertex function $\Gamma $ in the second order in the
screened interaction}
As an example, let us calculate the vertex function $\gamma _{\alpha \sigma
,\alpha \sigma }\left( p_{1};p_{2};k\right) .$ It is assumed in this paper that
the masses of the electrons and holes are the same and do not depend on the
subscript $\sigma .$ Then, the vertex function $\gamma _{\alpha \sigma
,\alpha \sigma }\left( p_{1};p_{2};k\right) $ does not depend on the $\sigma $ as well. It follows from Fig. \ref{Figure3} that
\begin{equation}
\gamma _{ee}=-\int \frac{d^{3}p}{\left( 2\pi \right) ^{3}}U_{ee}\left(
p\right) U_{ee}\left( k-p\right) \left[ G_{e}\left( p_{1}-p\right) \left(
G_{e}\left( p_{2}+p\right) +G_{e}\left( p_{2}+k-p\right) \right) \right] .
\label{eq8}
\end{equation}%
It is clear that the Green function $G_{e}\left( p\right) ~$ and the
screened interaction $U_{ee}\left( p\right) $ should be substituted into (%
\ref{eq8}) in the RPA. The mass operator $\Sigma _{\alpha \sigma }\left(
\omega ,\mathbf{p}\right) $ and the chemical potential $\mu _{\alpha }$
depend neither on the $\alpha $ nor on the $\sigma .$ The same takes
place for the Green function $G_{e}\left( p\right) =G\left( p\right) =\left(
i\omega +\mu -\Sigma \left( p\right) -\mathbf{p}^{2}/2\right) ^{-1}.$
According to Ref. \cite{JETLET}, $\mu =p_{F}^{2}/2+2\pi n/\nu +2\pi
nl-cn^{1/3}$and $\Sigma _{\alpha \sigma }\left( p\right) =2\pi n/\nu +2\pi
nl-cn^{1/3}$ in the RPA. Then, the Green function in the RPA reads
\begin{equation}
G\left( p\right) \approx \left( i\omega +p_{F}^{2}/2-\mathbf{p}^{2}/2\right)
^{-1},~p_{F}=2\pi ^{1/2}\left( n/\nu \right) ^{1/2}.  \label{eq8-1-1}
\end{equation}

If $l\ll a_{B}/\nu ,$ then the interaction $U_{ee}\left( p\right) $ in the
RPA reads\cite{JETLET}%
\begin{equation}
U^{\left( 0\right) }\left( p\right) =\frac{V\left( p\right) }{1-2V\left(
p\right) \Pi _{0}\left( p\right) }.  \label{eq8-1}
\end{equation}%
Here%
\begin{equation}
\Pi _{0}\left( p\right) =T/S\sum_{\mathbf{k},\varepsilon ,\sigma }G\left(
\varepsilon +\omega ,\mathbf{p}+\mathbf{k}\right) G\left( \omega ,\mathbf{k}%
\right),  \label{eq8-2}
\end{equation}%
where $G\left( p\right) $ is given by Eq. (\ref{eq8-1-1}).

We are interested in the behavior of the vertex function for the momenta
which obey condition (\ref{eq10}). The calculation of the integral like (\ref%
{eq8}) is considered in detail in Ref. \cite{JETLET}. The important feature of
these integrals is that the main contribution comes from the region $%
\left\vert \mathbf{p}\right\vert \sim $ $k_{0}\sim n^{1/3}$ and,
correspondingly, $\omega \sim k_{0}^{2}/2$ $\sim n^{2/3}.$ Let us remember
that we consider the values of the parameters such that the concentration $%
n\ll \nu ^{3}.$ Bearing this in mind, one has $k_{0}\gg p_{F}$ and $\omega
_{0}\gg \varepsilon _{F}.$ Then, the expression in the square brackets in (\ref{eq8}) is
reduced to these which coincides with the polarization operator asymptotics
\begin{equation}
\Pi _{0}\left( p\right) \approx -n\mathbf{p}^{2}/\left( \omega ^{2}+\left(
\mathbf{p}^{2}/2\right) ^{2}\right) ,\mbox{for }\left\vert \mathbf{p}%
\right\vert \sim p_{0}\gg p_{F},\mbox{and~}\omega \sim \omega _{0}\gg
\varepsilon _{F}  \label{eq7-1}
\end{equation}%
Substituting Eq.(\ref{eq7-1}) and (\ref{eq8-1}) into Eq. (\ref{eq8}), one
obtains
\begin{equation}
\gamma _{ee}\left( p_{1};p_{2};k\right) \approx \gamma =-\frac{1}{2}\int
\frac{d^{2}\mathbf{p}d\varepsilon }{\left( 2\pi \right) ^{3}}\left( \frac{%
V\left( p\right) \Pi _{0}\left( p\right) /n}{1-2n(V\left( p\right) \Pi
_{0}\left( p\right) /n)}\right) ^{2}.  \label{eq11}
\end{equation}%
The function $V\left( p\right) \Pi _{0}\left( p\right) /n$ does not depend
on the concentration $n.$ Then, the estimate of the integral (\ref{eq11})
gives
\begin{equation}
\gamma =-C/n^{2/3}.  \label{eq7-2}
\end{equation}%
Here $C$ is the factor of the order of unity. One can show that $\gamma
_{eh}\left( p_{1};p_{2};k\right) =\gamma _{ee}\left( p_{1};p_{2};k\right) ~$%
if $k_{0}l\ll 1.$
\begin{acknowledgements}
We are grateful to Yuri Kagan for valuable discussion. We acknowledge support from the Russian Fund For Basic Research (Grant 13-02-00472) and from the Ministry of Education and Science of Russian Federation (Project 8364)
\end{acknowledgements}

\pagebreak

\end{document}